\documentclass[aps,twocolumn,nofootinbib]{revtex4}
\usepackage{amsfonts}
\usepackage{amsmath}
\usepackage{amssymb}
\usepackage{graphicx}
\usepackage{comment}
\usepackage{color}
\usepackage{dialogue}
\usepackage{hyperref}
\providecommand{\U}[1]{\protect\rule{.1in}{.1in}}
\newtheorem{theorem}{Theorem}

\newtheorem{definition}[theorem]{Definition}

\begin{document}
\title{The status of determinism in proofs of the impossibility of a noncontextual model of quantum theory}

\author{Robert W. Spekkens}
\email{rspekkens@perimeterinstitute.ca}
\affiliation{Perimeter Institute for Theoretical Physics, 31 Caroline Street North, Waterloo, Ontario Canada N2L 2Y5}

\begin{abstract}
In order to claim that one has experimentally tested  whether a noncontextual ontological model could underlie certain measurement statistics in quantum theory, it is necessary to have a notion of noncontextuality that applies to unsharp measurements, 
i.e., those that can only be represented by positive operator-valued measures rather than projection-valued measures.
This is because any realistic measurement necessarily has some nonvanishing amount of noise 
and therefore never achieves the ideal of sharpness. 
Assuming a generalized notion of noncontextuality that applies to arbitrary experimental procedures, it is shown that the outcome of a measurement depends deterministically on the ontic state of the system being measured
if and only if the measurement is sharp.
Hence for every unsharp measurement, its outcome necessarily has an \emph{in}deterministic dependence on the ontic state.  We defend this proposal against alternatives.
In particular, we demonstrate why considerations parallel to Fine's theorem do not challenge this conclusion.

\end{abstract}
\date{Jan. 5, 2015}
\maketitle

\section{Introduction}

The Bell-Kochen-Specker theorem concerns the possibility of explaining the predictions of quantum theory in terms of a noncontextual ontological model~\cite{Bell,KochenSpecker}. In recent years, there has been some effort devoted to making the notion of a noncontextual
ontological model more operational~\cite{CabelloGarcia,Zeilinger,Larsson,Spe05}.  The goal of such efforts is to be able to decide, for any given operational theory (possibly distinct from quantum theory), whether it admits of such a model or not. Achieving this goal would allow one to assess the possibility of a noncontextual ontological model based solely on experimental data (regardless of whether or not this data is consistent with quantum theory): just as the observation of Bell
inequality violations~\cite{Belllocality} rule out a locally causal model of nature\footnote{More precisely, they rule out models that satisfy Bell's notion of local causality and certain other assumptions, such as the freedom of settings.}, irrespective of the truth of quantum theory, so too
would violations of operational \emph{noncontextuality inequalities} rule out a noncontextual ontological model of nature.

This article does not explicitly address the question of how to derive such inequalities.  Rather, it presents some preparatory work towards this goal.  We presume the correctness of operational quantum theory, but 
 we consider the predictions of quantum theory for experiments that do not satisfy the idealizations that are typically assumed in discussions of noncontextuality.  In particular, we relax the assumption
  that all the measurements in the experiment are projective.
 This assumption is an idealization because any realistic measurement procedure on a physical system is subject to some unavoidable noise, and consequently cannot be represented by a set of projectors on the Hilbert space of the system.  Rather, it must be represented by a positive operator valued measure (POVM), that is, a set of positive operators (called \emph{effects}) that sum to the identity operator.   

Recall that \emph{any} measurement procedure can be represented by a POVM.  In the special case where all of the effects in the POVM are projectors, we call the POVM a projector-valued measure and we call the measurement projective or \emph{sharp}.  If the POVM that represents a measurement is not a projector-valued measure, we call the measurement \emph{unsharp}.  The fact that no realistic measurement procedure is strictly free of noise implies that sharpness is an ideal that is never actually realized in any experiment.  Consequently, the question of how to represent \emph{unsharp} measurements in a noncontextual ontological model is critical if one wishes to compare the predictions of such a model to the predictions of quantum theory that actually arise in realistic experiments.

One proposal which has been made is that every unsharp measurement should be represented by a set of \emph{outcome-deterministic} response functions. 
What this means is that the ontological model specifies hidden variables which fix the outcome of each unsharp measurement (in a context-independent manner).  This assumption, of \emph{outcome determinism for all unsharp measurements}, will be abbreviated as ODUM.
It is explicit, for instance, in Ref.~\cite{Busch} (here $v(E)$ denotes the value assigned by the hidden variables to an effect $E$ appearing in a POVM):
\begin{quote}
An interpretation of valuations as truth value assignments would require the numbers $v(E)$ to be either 1 or 0, indicating the occurrence or nonoccurrence of an outcome associated with $E$. Valuations with this property are referred to as dispersion-free. 
\end{quote}
The assumption of ODUM is also made in Ref.~\cite{Cabello} (here, the discussion is of a particular set of unsharp quantum measurements):
\begin{quote}
Each [POVM in the set] contains eight positive-semidefinite operators whose sum is the identity. Therefore, a noncontextual hidden-variable theory must assign the answer yes to one and only one of these eight operators.
\end{quote}
In fact, this assumption has been made in almost every article that has considered noncontextuality for unsharp measurements~\cite{Busch,Cabello,Aravind, Methot,Renes,Zhang,Mancinska}.  

In this article, we will argue \emph{against} ODUM.

Adopting the operational notion of a noncontextual ontological model that the author proposed in Ref.~\cite{Spe05}, we will argue that unsharp measurements in quantum theory must be represented by outcome-\emph{in}deterministic response functions.  The representation is still noncontextual insofar as the \emph{probability} that one assigns to an outcome does not depend on the context.  The representation also remains highly constrained because sharp measurements will still need to be represented by a set of outcome-deterministic response functions (See Thm.~\ref{thm:main} below).\footnote{The fact that all noncontextual ontological models must represent sharp measurements outcome-deterministically implies in particular that the $\psi$-complete ontological model of quantum theory fails to be a noncontextual model.  The $\psi$-complete ontological model of quantum theory is the one wherein the ontic states are the pure quantum states (i.e. no hidden variables) and the response function associated to an effect is simply the conditional probability defined by the Born rule~\cite{HarriganSpekkens}.  This model is sometimes called the ``orthodox'' interpretation of quantum theory.  Given the use of the Born rule, the response functions in this model are outcome-indeterministic even for sharp measurements and therefore our result implies that the model cannot be noncontextual.  This inference is correct because the $\psi$-complete ontological model fails to be \emph{preparation noncontextual}, as explained in Sec. VIII.B of Ref.~\cite{Spe05}.} 
This fact about the representation of sharp measurements is relevant for realistic experiments, even though all the measurements in such experiments are unsharp.  This is because 
quantum theory dictates that a given unsharp measurement can be related in various ways to a set of idealized sharp measurements.
 For instance, it may be the case that the unsharp measurement realized in an experiment is a convex mixture of a set of idealized sharp measurements.  As another example, it is always possible to express an unsharp measurement on some system as the reduction of an idealized sharp measurement on a larger system.  Such relations typically imply that the response function for a given unsharp measurement is expressible in terms of the response functions for the idealized sharp measurements; the fact that the latter are required to be outcome-deterministic then imposes nontrivial constraints on the former.  

We begin by highlighting an intuitive (but mistaken) argument in favour of ODUM.
It is well-known in foundational circles that if one can find a local indeterministic ontological model for some set of correlations in a Bell-type experiment, then one can also find a local deterministic ontological model for the same set.  This was first noted by Fine~\cite{Fine}.  As such, if one can rule out local deterministic models for some set of correlations, then one has also ruled out local indeterministic models.  In other words, although one might have imagined that 
the class of correlations that can be explained by local indeterministic models is strictly larger than the class that can be explained by local deterministic models, it turns out that the classes of correlations that can be explained by the two sorts of models are precisely the same.

In discussions of noncontextuality, the role of determinism is less clear.  Kochen and Specker's notion of a noncontextual model explicitly incorporated the assumption that the outcomes of a measurement should be fixed deterministically by the hidden variables. 
One of the selling points of the operational definition of noncontextuality proposed in Ref.~\cite{Spe05}
is that it explicitly disentangles the notion of measurement noncontextuality from the assumption of outcome determinism.  
Furthermore, by using a notion of noncontextuality for \emph{preparations}, one can justify the assumption of outcome determinism for \emph{sharp} measurements~\cite{Spe05}. As such, the contradiction that is derived in the Kochen-Specker theorem forces us to reject \emph{some} assumption of noncontextuality.
However, there is nothing that warrants assuming outcome determinism for unsharp measurements.  
As such, any no-go theorem that assumes ODUM and derives a contradiction 
does not deserve to be called a proof of the impossibility of a noncontextual ontological model 
because in the face of a contradiction, one can always assume that the faulty assumption was that of outcome determinism rather than that of noncontextuality.

Nonetheless, one might wonder whether a version of Fine's theorem applies to noncontextual models just as it does to locally-causal models. That is, one might wonder whether arguments analogous to those of Fine could establish that one can rule out noncontextual ontological models that are outcome-indeterministic for unsharp measurements if and only if one can also rule out those that are outcome-deterministic for these measurements.  If this were so, then there would be no loss of generality in assuming ODUM, and consequently no good reason not to do so.

In fact, in Ref.~\cite{SimonBruknerZeilinger2001}, this is the reason given for not considering outcome-indeterministic models:
\begin{quote}
[We] do not discuss stochastic hidden variable theories explicitly.  This does not limit the generality of the results derived because the existence of a stochastic local or noncontextual hidden variable model for a given physical system implies that also an underlying deterministic local or noncontextual model can be constructed which reproduces the probabilities of the stochastic model.  Therefore, e.g., ruling out all possible noncontextual deterministic hidden variable models implies ruling out all possible noncontextual stochastic models as well.
\end{quote}

In this article, we will explain why this intuition is mistaken, that is, why Fine's theorem does not extend to noncontextual models in the manner that would be required to justify ODUM.  The explanation, in essence, is the following: while it is true that one can always model a measurement that  depends indeterministically on the ontic state of the system by one that depends deterministically on the ontic state of a larger system (including, for instance, hidden variables in the apparatus), nevertheless, under such a change of representation one finds that measurement events that differ only by a choice of context are mapped to measurement events that differ in their operational statistics and consequently one loses the warrant to apply the assumption of noncontextuality.  

We also provide a number of additional arguments against outcome-deterministic representations of unsharp measurements in noncontextual models and against variants of these arguments that seek to limit the sorts of unsharp measurements to which the assumption of noncontextuality can be applied.

In addition, we discuss some of the consequences of our result that measurements are represented outcome-deterministically if and only if they are sharp.  For instance, the assumption of measurement noncontextuality implies the existence of a valuation that assigns to each effect a probability (independently of the POVM in which that effect appears), and our results imply that the value assigned to an effect must be one of its eigenvalues.
This observation allows us to provide significantly simplified versions of two proofs of the impossibility of a noncontextual ontological model of operational quantum theory which were first presented in Ref.~\cite{Spe05} and which make explicit use of the assumption of noncontextuality for unsharp measurements.

The article is organized as follows.  In Sec.~\ref{Preliminaries}, we provide definitions of the relevant background concepts for this article: operational theories, ontological models, noncontextuality, and outcome determinism.  Sec.~\ref{sec:compellingargument} describes the intuition in favour of ODUM (along the lines of Fine's theorem) and why it is mistaken.  Additional supporting material for this section is provided in the appendices.   Sec.~\ref{sec:OIUM} presents the proof that within a noncontextual ontological model, a measurement is represented outcome-deterministically if and only if it is sharp.  We also discuss which predictions of operational quantum theory are required for the proof and the significance of the result.  In Sec.~\ref{sec:probassignments}, we discuss how post-processing of a measurement is represented in an ontological model and draw out the consequences for noncontextual representations of sharp and unsharp measurements.  In particular, we formulate the constraints on noncontextual representations of measurements in terms of probability-assignments to effects.  Sec.~\ref{sec:coinflipPOVM} presents a few more arguments against ODUM, in the form of a dialogue between the author and an imaginary proponent of ODUM.  Finally, in Sec.~\ref{sec:discussion}, we use the constraints on noncontextual representations of measurements in terms of probability assignments to effects (introduced in Sec.~\ref{sec:probassignments}) to provide simple proofs of no-go theorems for noncontextuality that make explicit use of unsharp measurements.  

\section{Generalized noncontextuality and outcome-determinism}\label{Preliminaries}

An \emph{operational theory} is a triple $\left(  \mathcal{P},\mathcal{M}
,p\right)$ where 
$\mathcal{P}$ is a set of preparation procedures, an element of which is denoted by $P$, 
where $\mathcal{M}$ is a set of measurement procedures, an element of which is denoted by $M$ and for which the corresponding set of outcomes (assumed here to be discrete) is denoted by $\mathcal{K}_{M}$,
and where $p:\mathcal{P}\times\mathcal{M}\rightarrow [0,1]^{\mathcal{K}_M}::(P,M) \mapsto p(k|M,P)$ is a
nonnegative function satisfying $\sum_{k\in\mathcal{K}_{M}}p(k|M,P)=1$ for all
$M\in\mathcal{M}$ and all $P\in\mathcal{P}.$ \ $p(k|M,P)$ specifies the
probability of obtaining outcome $k$ in a measurement of $M$ following a
preparation $P.$  In brief, an operational theory specifies the possible
measurements and preparations and the statistics one obtains over the outcomes
of any measurement given any preparation.

An \emph{ontological model} for an operational theory $\left(  \mathcal{P}%
,\mathcal{M},p\right)  $ is a triple $\left(  \Lambda,\mu,\xi\right)  $
where $\Lambda$ is a discrete set\footnote{More generally, an ontological model may be defined such that $\Lambda$ is a continuous space, in which case $\mu:\mathcal{P}\times\Lambda\rightarrow
\mathbb{R}_+$ is a probability density and sums over $\lambda$ are replaced by integrals.}, where $\mu:\mathcal{P}\times\Lambda\rightarrow[0,1] 
:: (P,\lambda) \mapsto \mu(\lambda|P)$ is a nonnegative function satisfying 
$\sum_{\lambda\in\Lambda}
\mu\left(  \lambda|P\right)  =1$ for all
$P\in\mathcal{P},$ where $\xi:\mathcal{M}\times\Lambda\rightarrow
\lbrack0,1]^{\mathcal{K}_M}:: (M,\lambda) \mapsto \xi (k | \lambda, M)$ is a nonnegative vector function satisfying $\sum_{k\in\mathcal{K}_{M}}%
\xi \left(k | \lambda, M\right)  =1$ for all $\lambda\in\Lambda$ 
and for all $M  \in\mathcal{M},$ 
and where
\begin{equation}\label{eq:empiricaladequacy}
\sum_{\lambda\in\Lambda}
\mu\left(  \lambda|P\right)  \xi \left(k | \lambda, M\right)  =p\left(  k|M,P\right) ,
\end{equation}
for all $M \in \mathcal{M}$ and $P \in \mathcal{P}$.
The interpretation of these mathematical expressions is as
follows: $\lambda \in \Lambda$ denotes an \emph{ontic state}, that is, a complete specification of all the physical properties of the system (including a valuation of any hidden variables), and $\Lambda$ is the space of possible ontic states for the system; $\mu\left(
\lambda|P\right)  $ is the probability that the system is in ontic state
$\lambda$ given that it was prepared according to preparation $P$, and
$\xi \left(k | \lambda, M\right)$ is the probability that measurement $M$
yields outcome $k$ given that the system which was fed into the measurement device had ontic state $\lambda.$  The set $\{ \xi \left(k | \lambda, M\right) : k \in\mathcal{K}_M \}$, considered as functions over $\Lambda$, will be called the set of \emph{response functions} associated with $M$.
Equation~\eqref{eq:empiricaladequacy} asserts that the overall probability of outcome $k$ given
preparation $P$ and measurement $M$ specified by the operational theory is recovered by the ontological model.

According to the proposal of Ref.~\cite{Spe05}, noncontextuality is a
property of an ontological model of an operational theory. A distinction is also made between two sorts of noncontextuality, one which pertains to the
representation of preparations and the other of which pertains to the representation of
measurements.

\begin{definition}[noncontextuality]\label{def:NC}
An ontological model $\left(  \Lambda,\mu,\xi\right) $ of an operational theory $\left(  \mathcal{P},\mathcal{M}
,p\right)$  is said to be \emph{preparation noncontextual} if
the following implication holds: for any pair of preparation procedures $P,P' \in \mathcal{P}$,
\begin{align}
\textrm{if  }p(k|P,M)&=p(k|P^{\prime},M)\;
\forall M \in \mathcal{M} \nonumber \\
\textrm{then  }\mu(\lambda|P)&=\mu(\lambda|P^{\prime})\; \forall \lambda \in \Lambda. \label{eq:PNC}
\end{align}
In other words, preparation noncontextuality is the assumption  that if two preparations yield the same statistics for all possible
measurements then they are represented equivalently in the ontological model.

Similarly, an ontological model of an operational theory is said to be \emph{measurement noncontextual} if the following implication holds: for any pair of measurement procedures $M,M' \in \mathcal{M}$,
\begin{align}
\textrm{if  }p(k|P,M)&=p(k|P,M^{\prime})
\;\;\forall P\in \mathcal{P} \nonumber \\
\textrm{then  }\xi \left(k | \lambda, M\right)&= \xi \left(k | \lambda, M^{\prime}\right)
\;\; \forall \lambda \in \Lambda. \label{eq:MNC}
\end{align}
In other words, measurement noncontextuality is the assumption that if two measurements have the same statistics for all possible
preparations then they are represented equivalently in the ontological model.
\end{definition}

It is the notion of measurement noncontextuality that will be of particular interest to us here.  

In order to understand this notion better, it is useful to adopt the following terminological conventions. 
In an operational theory, the \emph{operational statistics }of a measurement $M$ are the probability distributions over outcomes for all possible preparation procedures$,$ that is, the set 
$\left\{  \left( p\left(  k|M,P\right) : k \in \mathcal{K}_M \right) : P \in \mathcal{P} \right\}.$
If two measurements $M$ and $M'$ have the same operational statistics, then they are said to be \emph{statistically indistinguishable relative to all preparations}.
In an ontological model of an operational theory, the \emph{ontological statistics }of a measurement $M$ are the probability distributions over outcomes for all possible ontic states $\lambda,$ that is, the set $\left\{  \left( \xi\left(  k|M,\lambda\right)  : k \in \mathcal{K}_M \right) : \lambda \in \Lambda \right\}.$  
If two measurements $M$ and $M'$ have the same ontological statistics, then they are said to be \emph{statistically indistinguishable relative to all ontic states}.
Given this terminology,  the assumption of measurement noncontextuality can be summarized as simply:
\begin{quote}
An ontological model of an operational theory is measurement noncontextual if measurements that are statistically indistinguishable relative to all preparations are statistically indistinguishable relative to all ontic states.
\end{quote}

We now provide a definition of the notion of outcome determinism, the central topic of this article.  We here make use of the notion of a \emph{measurement event}, which refers to the pair consisting of a measurement and an outcome of that measurement. 
\begin{definition}[outcome determinism]
A response function $\xi \left(k | \lambda, M\right)$ representing a measurement event (outcome $k$ of measurement $M$) is said to be \emph{outcome-deterministic }if
\begin{equation}
\xi \left(k | \lambda, M\right)  \in\left\{  0,1\right\}  \;  \forall \lambda\in\Lambda,
\end{equation}
that is, if the probability of obtaining the outcome in question, conditioned on the ontic state $\lambda$, is either $0$ or $1$.  Otherwise, it is said to be \emph{outcome-indeterministic.}  
A measurement $M$ associated to a set of response functions $\{ \xi \left(k | \lambda, M\right) : k \in \mathcal{K}_M \}$ is said to be represented \emph{outcome-deterministically }if every response function in the set is outcome-deterministic.  Otherwise, it is said to be represented outcome-indeterministically. 
\end{definition}

We will also say that we have outcome determinism (respectively outcome indeterminism) \emph{for some class of measurements} if every measurement in the class is represented outcome-deterministically  (respectively outcome-indeterministically).
\footnote{There is a subtlety in our choice of terminology which might lead to confusion.   For a set of response functions to be considered outcome-indeterministic, it is sufficient for a single response function in the set to be outcome-indeterministic.  Hence, every set of response functions is either outcome-deterministic or outcome-indeterministic.  On the other hand, for a given class of measurements to be considered outcome-indeterministic, it is necessary for \emph{every} measurement in the class to be outcome-indeterministic.  Consequently, we might have neither outcome-determinism nor outcome-indeterminism for a given class.} 

 We can therefore enumerate the three possibilities for how ontological models of quantum theory might treat sharp measurements:
 \begin{itemize}
\item[(a)] Outcome determinism for all sharp measurements
\item[(b)] Outcome indeterminism for all sharp measurements
\item[(c)] Outcome determinism for some sharp measurements and outcome indeterminism for others
\end{itemize}
and the three possibilities for how they might treat unsharp measurements
\begin{itemize}
\item[(a$^\prime$)] Outcome determinism for all unsharp measurements (abbreviated ODUM)
\item[(b$^\prime$)] Outcome indeterminism for all unsharp measurements
\item[(c$'$)] Outcome determinism for some unsharp measurements and outcome indeterminism for others
\end{itemize}

It turns out that for ontological models of operational quantum theory,  a noncontextual representation of sharp measurements must be outcome-deterministic, i.e., option (a), and a noncontextual representation of unsharp measurements must be outcome-indeterministic, i.e., option (b$'$).
Indeed, this is the main positive result of our article, which we summarize in the following theorem.

\begin{theorem}\label{thm:main}
In order to reproduce certain simple features of operational quantum theory (specifically, the features $P1$ and $P2$ outlined in Sec.~\ref{sec:OIUM}) any ontological model that is noncontextual in the sense of definition \ref{def:NC} must be such that a measurement is represented outcome-deterministically if and only if it is sharp.
\end{theorem}
The proof of this theorem, as well as a discussion of its signficance, is provided in Sec.~\ref{sec:OIUM}.  
First, however, we consider the question raised in the introduction concerning whether considerations parallel to Fine's theorem can justify ODUM.

\section{The intuition in favour of ODUM and why it is mistaken}\label{sec:compellingargument}

The most compelling (albeit incorrect) argument in favour of ODUM has the
following structure.

\begin{quote}
Premiss 1: If two measurements are statistically indistinguishable for all
preparations, then they are represented in the ontological model by the same set of response functions (this
is the assumption of measurement noncontextuality).

Premiss 2: Every measurement represented in the ontological model by an outcome-indeterministic
set of response functions on the system can also be represented by an
outcome-deterministic set of response functions on a composite of the system and an ancilla.

Purported conclusion: If two measurements are statistically
indistinguishable for all preparations, then they can be represented in the ontological model by the same
outcome-deterministic set of response functions.
\end{quote}

The flaw in the argument is not in the premisses; if these are clarified
appropriately, they are correct. \ It is that the conclusion does not follow
from the premisses. \ 

To get a feeling for why one ought to be suspicious of the implication,
consider the following analogous argument.

\begin{quote}
Premiss: If two measurements are statistically indistinguishable for all
preparations, then they are represented in quantum theory by the same POVM.

Premiss: Every measurement represented in quantum theory by a POVM on the system can be
represented by a projector-valued measure on a composite of the system and an ancilla (this is Naimark's theorem~\cite{Peres}).

Purported conclusion: If two measurements are statistically indistinguishable for all preparations,
then they can be represented in quantum theory by the same projector-valued measure.
\end{quote}

Here it is clear that the two premisses are true but the conclusion is false.
Two measurements procedures on system+ancilla might be represented by
different projector-valued measures, while they nonetheless reduce to the same POVM on the system and are therefore statistically
indistinguishable relative to all preparations of the system. This analogy is meant only to give a sense of
where the problem lies.

In the following, we will make an assumption about how the ontological model treats composite systems, namely, that the ontic state space of the composite is the Cartesian product of the ontic state spaces of the components.  For instance, if system $s$ has ontic state space $\Lambda_s$ and ancilla $a$ has ontic state space $\Lambda_a$ then the composite has ontic state space $\Lambda_{sa}=\Lambda_s \times \Lambda_a$.  This assumption has been called \emph{separability}~\cite{HarriganSpekkens} or \emph{kinematical locality}~\cite{SpekkensFQXI}.

We argue that the premisses and the conclusion of the argument in favour of ODUM should be made precise as follows:

\begin{quote}
Premiss $1^{\prime}$: If two measurements on a system $s$ are statistically
indistinguishable relative to all preparations on $s$, then they are
represented in the ontological model by the same set of response functions on $\Lambda_{s}$
(this is the assumption of measurement noncontextuality applied to system
$s$).

Premiss $2^{\prime}$: Every measurement on a system $s$ that is represented in the ontological model by
an outcome-indeterministic set of response functions on $\Lambda_{s}$ can also be
understood as the effective measurement arising from a preparation of an
ancilla $a$ and a measurement on system+ancilla $sa$ where the latter is
represented by an outcome-deterministic set of response functions on $\Lambda
_{s}\times\Lambda_{a}$ (the measurement on $sa$ is said to be the \emph{extension} of
the one on $s$).

Purported conclusion: If two measurements on a system $s$ are
statistically indistinguishable relative to all preparations on $s$, then the
two measurements on system+ancilla $sa$ that extend these can be represented
ontologically by the same outcome-deterministic set of response function on
$\Lambda_{s}\times\Lambda_{a}.$ \ 
\end{quote}

An explicit argument for premiss $2'$ is provided in appendix~\ref{app:ontext}.\footnote{Harrigan and Rudolph have also previously discussed ontological models that assign ontic state spaces to the measurement apparatus~\cite{HarriganRudolph}.} Our task now is to show why the conclusion does not follow. It suffices to note the following.

\begin{quote}\label{prop:operationalextension}
Claim: If two measurements on a system $s$ are statistically indistinguishable
relative to all preparations on $s$, it may nonetheless be the case that the
two measurements on system+ancilla $sa$ that are extensions of these are
statistically distinguishable for some preparation on $sa.$
\end{quote}

This claim is seen to be true simply by noting that a given POVM can have
two distinct Naimark extensions, that is, there are multiple choices of
projector-valued measure on $sa$ that yield the same\ POVM on $s.$  
Although this fact is well-known, we provide an explicit example in appendix~\ref{app:Naimark}.
It  blocks the purported conclusion because if two measurements on system+ancilla $sa$
are statistically distinguishable for some preparation on $sa,$ then they must
be represented ontologically by different response functions on $\Lambda
_{s}\times\Lambda_{a}$. 

The conclusion we draw from all of this is that one can model a pair of
measurements that are statistically indistinguishable for all preparations on some system by either:
\begin{itemize}
\item[(i)] the \emph{same}  outcome-\emph{indeterministic} set of response functions on the system,
\item[(ii)] \emph{different} outcome-\emph{deterministic} sets of response functions on system+ancilla.
\end{itemize}
So, it is true that one can always represent a measurement by an outcome-deterministic
set of response function if one wishes; this is done by incorporating other degrees
of freedom (for instance those of the apparatus) in one's description.
\ However, in moving to such a representation, one loses the warrant to apply the
assumption of measurement noncontextuality.   In particular, although two measurements on a system may be represented by the same POVM in quantum theory, this is not enough to justify (via the assumption of measurement noncontextuality) that they should be represented by the same set of response functions on system+ancilla in the ontological model. The latter conclusion would only be justified if the two measurements were represented by the same projector-valued measure on system+ancilla and this is generally not the case.  

We can investigate the possibility of a noncontextual ontological model using either of conventions (i) or (ii).
It is useful to consider how one models an unsharp measurements on system $s$, denoted $M_s$, within each convention.

Under convention (i),  we confine our attention to the ontic state space of system $s$, $\Lambda_s$, we assume outcome indeterminism,
\begin{equation}
\xi \left(k | \lambda, M_s \right)  \in [0,1] \;\; \forall \lambda \in \Lambda_s \text{,}
\end{equation}
and we represent the assumption of measurement noncontextuality as
\begin{align}
\textrm{If  }p(k|P_s,M_s)&=p(k|P_s,M_s^{\prime})
\;\;\forall P_s\in \mathcal{P}_s \nonumber \\
\textrm{then  }\xi \left(k | \lambda, M_s\right)&= \xi \left(k | \lambda, M_s^{\prime}\right)
\;\; \forall \lambda \in \Lambda_s. \label{eq:MNC}
\end{align}

Under convention (ii), on the other hand, we expand the ontic state space from that of the system to that of system+ancilla, $\Lambda_{s}\times\Lambda_{a},$  such that we can assume outcome determinism,%
\begin{equation}
\xi \left( k| \lambda_{s},\lambda_{a}, M_{sa}\right)  \in\{0,1\} \;\; \forall (\lambda_s, \lambda_a) \in \Lambda_a \times \Lambda_a \text{,}%
\end{equation}
but in this case we must represent the assumption of measurement noncontextuality as
\begin{align}\label{eq:newMNC}
\text{if }&p\left(  k|P_{s}, M_{s}\right)   = p\left(  k|P_{s}, M_{s}^{\prime}
\right)  
\;\forall P_s\in \mathcal{P}_s
\text{, }\nonumber \\
 \text{then }&\sum_{\lambda_{a}}\mu\left(  \lambda_{a}| P_{a}\right)
\xi \left(k|  \lambda_{s},\lambda_{a}, M_{sa}\right) \nonumber \\
& =\sum_{\lambda_{a}}\mu \left(  \lambda_{a}| P_{a}^{\prime}\right)
\xi \left( k|\lambda_{s},\lambda_{a}, M_{sa}^{\prime}\right)
\;\forall \lambda_s \in \Lambda_s .
\end{align}
In Eq.~\eqref{eq:newMNC}, the expressions on either side of the equality in the consequent 
are the effective response functions on the system.
If we think of Eq.~\eqref{eq:newMNC} as a constraint on the outcome-deterministic response functions $\xi\left( k | \lambda_{s},\lambda_{a} , M_{sa} \right)$ and $\xi \left(  k|\lambda_{s},\lambda_{a}, M_{sa}^{\prime}\right)$, it is not simply a constraint of equality of those response functions.  The more cumbersome nature of this constraint is the price to pay for insisting on outcome-deterministic representations.

In appendix \ref{app:exampleontext} we provide an explicit example of how, for a particular pair of measurements on the system, each measurement can be associated with a set of response functions that are outcome-deterministic on an extended system---via the scheme of appendix~\ref{app:ontext}---but the two sets of response functions fail to be equivalent.

These two approaches to providing a noncontextual ontological model are analogous to two approaches to operational quantum theory known as the ``church of the larger Hilbert space'' and the ``church of the smaller Hilbert space''\footnote{Terms coined by John Smolin and Matt Leifer respectively.}.  In the former, preparations, measurements, and transformations on a system are represented respectively by pure states, projector-valued measures, and unitary maps on system+ancilla, while in the latter, they are represented by mixed states, POVMs and completely positive trace-preserving maps on the system alone.  Just as this choice is conventional---either can do justice to the experimental statistics---so too is the choice of whether to posit an ontological model on the system alone or on system+ancilla.  Indeed, it is appropriate to refer to convention (ii) as the ``church of the larger ontic state space'' and convention (i) as ``the church of the smaller ontic state space''.

Conventions (i) and (ii) encode one and the same notion of noncontextuality for unsharp measurements.  The point is that this notion is distinct from the one that is
recommended in Refs.~\cite{Busch,Cabello,Aravind, Methot,Renes,Zhang,Mancinska}.

We shall adopt convention (i), i.e. the church of the smaller ontic state space, for the rest of the article. 

\section{Proving that outcome indeterminism holds for all and only unsharp measurements}
\label{sec:OIUM}

We will now prove Theorem~\ref{thm:main}, which asserts that, given certain features of operational quantum theory,
it follows that in a noncontextual ontological model (using the notion of noncontextuality in definition \ref{def:NC}) the set of response functions  associated with a measurement is outcome-deterministic if and only if the measurement is sharp, or equivalently, that it is outcome-indeterministic if and only if the measurement is unsharp. 

We divide the proof into two halves:
\begin{itemize}
\item[(a)]  if a measurement is sharp then it is represented outcome-deterministically in a noncontextual ontological model,  
\end{itemize}
as well as the converse of this implication, which in its contrapositive form is
\begin{itemize}
\item[(b)]   if a measurement is unsharp then it is represented outcome-\emph{in}deterministically in a noncontextual ontological model,  
\end{itemize}

For the proof of (a)  we refer the reader to Sec.~VIII.A of Ref.~\cite{Spe05}. As noted there, one needs the generalized notion of noncontextuality from definition \ref{def:NC} to infer outcome determinism for sharp measurements, specifically, one needs the assumption of \emph{preparation noncontextuality}; the assumption of measurement noncontextuality alone is insufficient to derive the result. 

The features of operational quantum theory that are used in the proof of (a) can be summarized as follows (as one can infer from the details of the proof, as described in Ref.~\cite{Spe05}):
\begin{itemize}
\item[P1] For every sharp measurement $M$, and for each outcome $k$ of that measurement, there is a preparation procedure $P_{M,k}$ that makes that outcome certain to occur, that is, $p(k|M,P_{M,k'})=\delta_{k,k'}$.  Furthermore, for any two sharp measurements $M$ and $M'$, the uniform mixture over $k$ of the $P_{M,k}$ and the uniform mixture over $k$ of the $P_{M',k}$ are statistically indistinguishable relative to all measurements.
\end{itemize}

The proof of (b) is a novel contribution of this article.  The only sort of noncontextuality that is relevant here is measurement noncontextuality.

In the course of the proof, we will make critical use of the following fact:
If it can be shown that \emph{one particular} measurement procedure realizing a given POVM must be represented outcome-indeterministically, then  in a noncontextual ontological model it follows that \emph{all} measurement procedures realizing that POVM must be represented outcome-indeterministically.  The reason is that if the two measurement procedures realize the same POVM, then according to operational quantum theory, they are statistically indistinguishable relative to all preparations, and the assumption of measurement noncontextuality then implies that they must be represented by the same set of response functions.

To begin, it is useful to note how post-processing of measurements are represented in an ontological model.  These constraints have nothing to do with the assumption of noncontextuality.  They are constraints on \emph{any} ontological model, contextual or noncontextual. 

Suppose a measurement $M^{\prime}$ is defined in terms of post-processing of another measurement $M$ as follows. \ 
\begin{quote}
The procedure $M^{\prime}$ : Implement $M$ and upon obtaining outcome $k,$
sample a random variable $j$ from a conditional probability distribution $s(j|k)$.
Finally, output
$j$ as the outcome of the effective measurement.
\end{quote}
Now consider how these measurements must be represented in an ontological model. If outcome $k$ of measurement $M$ is represented by the
response function $\xi(k|\lambda,M),$ the outcome $j$ of measurement $M^{\prime
}$ is represented by the response function
\begin{equation}\label{eq:postprocessing}
\xi \left(  j|\lambda,M' \right)  =\sum_{k}s\left(  j|k\right)  \xi (k|\lambda,M).
\end{equation}
This just follows from probability theory: we must sum the probability of all the ways of getting outcome $j$ in measurement $M'$.  We therefore must take the sum over $k$ of the probability of getting outcome $k$ in measurement $M$ and of then getting outcome $j$ in the sampling, and this latter probability is simply the product of the probability of getting outcome $k$ in $M$ and the probability of getting $j$ given $k$. 

Note that coarse-graining of measurement outcomes is a special case of post-processing.  For instance, if one wishes to coarse-grain all outcomes $k$ in some subset $S$ to a single outcome $j=j_0$, one simply chooses the conditional such that $s\left(  j_0|k\right)=1$ for all $k\in S$.  In this case,  
\begin{equation}
\label{eq:coarsegraining}
\xi \left(  j_0|\lambda,M' \right)  = \sum_{k\in S} \xi (k|\lambda,M).
\end{equation}
Therefore coarse-graining of outcomes in the operational theory is represented by coarse-graining of the corresponding response functions.  

Now consider the operational equivalence class of measurement events that are associated with a particular effect $E$. 
Denote the spectral resolution of $E$ by $E=\sum_i s_i \Pi_i$ where the $\Pi_i$ are projectors satisfying $\sum_i \Pi_i = I$.  
We can use the spectral resolution of $E$ to build up a measurement $M'$ that has an outcome that is in the equivalence class associated with $E$.  
First, let $M$ denote a sharp measurement associated with the  projector-valued measure $\{\Pi_i\}$. Next, define $M'$ to be a post-processing of $M$ wherein the conditional probability $s(j|i)$ is chosen so that for some value of $j$, denoted $j_0$, we have $s(j_0|i) =s_i$.  The $j_0$ outcome of measurement $M'$ will then be associated with the effect $\sum_i s_i \Pi_i$, which is $E$.  Because of how post-processing is represented in an ontological model (see Eq.~\eqref{eq:postprocessing}), we have 
\begin{equation}\label{eq:jjj}
\xi(j_0|\lambda,M')=\sum_i s_i \xi(i|\lambda,M).
\end{equation}
As emphasized above, the assumption of measurement noncontextuality then implies that for \emph{every} measurement event associated with effect $E$, not just the event corresponding to the $j_0$ outcome of measurement $M'$, the response function for this event is the same.  We denote it by $\xi_E(\lambda)$.  Denoting the response function for the operational equivalence class of the projector $\Pi_i$ by $\xi_{\Pi_i}(\lambda)$, we infer from Eq.~\eqref{eq:jjj} that
\begin{equation}\label{eq:proofthm}
\xi_E(\lambda)=\sum_i s_i \xi_{\Pi_i}(\lambda).
\end{equation}

We are now in a position to demonstrate that all unsharp measurements must be represented outcome-indeterministically.  If a quantum measurement is unsharp, then at least one of its outcomes is associated with an effect that is not a projector.  Call this effect $E$ and denote its spectral resolution by  $E=\sum_i s_i \Pi_i$ as before, so that its response function is given by Eq.~\eqref{eq:proofthm}.  The fact that $E$ is not a projector implies that one of its eigenvalues, say $s_{i_0}$, is such that $0 < s_{i_0} < 1$.  Given that $\{ \Pi_i \}$ is itself a POVM, 
it follows that $\sum_i \xi_{\Pi_i}(\lambda)=1$ in the ontological model (because for every $\lambda$, some outcome of the measurement associated with $\{ \Pi_i \}$  must occur).   But $s_{i_0} < 1$ and $\sum_i \xi_{\Pi_i}(\lambda)=1$ together imply that $\sum_i s_i \xi_{\Pi_i}(\lambda) < 1$.  Finally, because there exist quantum states that assign a nonzero probability to the $i_0$ outcome of the $\{ \Pi_i \}$ measurement, it follows that the response function associated with $i_0$ must have nontrivial support on the ontic state space.  Hence, for every $\lambda$ in this support, we have $0< \sum_i s_i \xi_{\Pi_i}(\lambda) < 1$, which implies that $0< \xi_{E}(\lambda) < 1$.  So, $\xi_E(\lambda)$ is outcome-indeterministic.

It is straightforward to verify that the features of operational quantum theory that are used in the proof of (b) are:
\begin{itemize}
\item[P2] For every unsharp measurement $M'$, the measurement event associated to any given outcome of $M'$
can be realized by a post-processing of some sharp measurement $M$ where the post-processing is intrinsically probabilistic rather than deterministic.
In other words, if the outcomes of $M$ are labelled by $i$, then there exists at least one such value, $i_0$, such that the conditional probability $s_{i_0}$ of obtaining the distinguished outcome of $M'$ given that outcome $i_0$ was obtained for $M$ satisfies $0 < s_{i_0}<1$.
\end{itemize}

This concludes the proof of theorem~\ref{thm:main}.

\subsection{The significance of Theorem~\ref{thm:main}}

It is worth commenting on how one should interpret Thm.~\ref{thm:main} given that it is known that operational quantum theory does \emph{not} admit of an ontological model that is noncontextual in the sense of definition~\ref{def:NC}.  In particular, in Ref.~\cite{Spe05}, it was shown that preparation noncontextuality \emph{alone} implies a contradiction with the predictions of operational quantum theory.  As such, one might wonder what is the use of characterizing how measurements must be represented in noncontextual ontological models of operational quantum theory, as Thm.~\ref{thm:main} does, when it is known that there are no such models.\footnote{This question was posed by a referee of this article.}  The answer is that Thm.~\ref{thm:main} is a tool for devising novel proofs of the impossibility of a noncontextual ontological model of quantum theory, in particular, proofs that make nontrivial use of the assumption of measurement noncontextuality.  Indeed, two examples of such proofs are provided in Sec.~\ref{sec:discussion}.

There are many benefits to having multiple different proofs of the impossibility of a noncontextual ontological model of operational quantum theory.  After all, the point of studying noncontextual ontological models is not merely to rule them out, but to get a more complete picture of the sense in which operational quantum theory differs from a classical theory.  

It is perhaps easiest to appreciate this point by considering its analogue in the context of proofs of the inconsistency of operational quantum theory and Bell's assumption of local causality.  No one can deny that much has been learned by exploring the possible paths that such inferences might take, such as the proof provided by Hardy~\cite{Hardy} or the proof provided by Greenberger Horne and Zeilinger~\cite{GHZ}.  
Just as the different proofs of Bell's theorem vary in interesting ways, so too do the different proofs of the impossibility of a noncontextual ontological model of operational quantum theory.

Also, many of the information-processing advantages of quantum theory can be proven to be connected to the impossibility of a noncontextual model.  The cryptographic task of \emph{parity-oblivious multiplexing}, a kind of random access code, is an example~\cite{ParityObliviousMulitplexing}.  It has also recently been shown that quantum contextuality is a resource in the magic state distilation model of fault-tolerant quantum computation~\cite{magic}.   
Understanding the various different logical paths from the assumption of noncontextuality to a contradiction is important for determining which quantum information-processing tasks might be powered by contextuality. 

It is also worth noting that the particular features of operational quantum theory that are needed to prove Thm.~\ref{thm:main}, namely $P1$ and $P2$, are not by themselves sufficient to derive a contradiction with the assumption of a noncontextual ontological model.  One requires additional features of operational quantum theory to obtain the contradiction.
To see that this is the case, 
it suffices to show that there are subtheories of operational quantum theory---that is, subsets of the full set of preparations and measurements---that satisfy $P1$ and $P2$ and admit of a noncontextual ontological model. 

One example is the stabilizer subtheory for qutrits~\cite{Gross}.  This is the set of experiments on collections of three-level quantum systems (qutrits) of the following form: the preparations are those associated with stabilizer states (density operators that are normalized projectors onto an eigenspace of a commuting set of products of generalized Pauli operators); the sharp measurements are Pauli measurements (those associated with a commuting set of products of generalized Pauli operators), and the unsharp measurements are those which have a Naimark extension wherein the ancilla is prepared in a stabilizer state and subjected to a Pauli measurement.  

Another example is Gaussian quantum mechanics~\cite{Liouville}.  This is the set of experiments on collections of continuous-variable systems of the following form: the preparations are those associated with Gaussian states (density operators with Gaussian Wigner representation); the sharp measurements are those associated with projector-valued measures every element of which has a Gaussian Wigner representation; and the unsharp measurements are those which have a Naimark extension wherein the ancilla is prepared in a Gaussian state and subjected to a sharp Gaussian measurement. 

Both of these subtheories of operational quantum theory satisfy $P1$ and $P2$ and therefore have the features that are necessary for proving Thm.~\ref{thm:main}.  However, unlike the full quantum theory, they \emph{do} admit of an ontological model that is noncontextual in the sense of definition~\ref{def:NC}.  This follows from the fact that the preparations and measurements in these subtheories have nonnegative Wigner representations (in the stabilizer subtheory, it is the discrete Wigner representation of Gross~\cite{Gross}) and the fact that a nonnegative quasiprobability representation yields a noncontextual ontological model, as shown in Ref.~\cite{Spe08}. It follows from Thm.~\ref{thm:main} that in these ontological models, measurements are represented outcome-deterministically if and only if they are sharp. By inspecting the noncontextual ontological models that the Wigner representation defines, one can verify that this is indeed the case.

\section{Constraints on noncontextual probability-assignments over effects}\label{sec:probassignments}

In quantum theory, two measurements are statistically indistinguishable relative to all preparations if they are represented by the same POVM.  The POVMs, therefore, describe the operational equivalence classes of measurements. 
It follows that if a measurement $M$ is represented by the POVM $\{ E_k \}$, then measurement noncontextuality implies that the set of response functions representing $M$ can be labelled by the POVM alone, i.e. no other details of the measurement procedure need to be specified,
\begin{equation}\label{eq:povmdep}
\xi(k|\lambda,M)=\xi(k|\lambda,\{ E_k \}).
\end{equation}

In fact, measurement noncontextuality implies a further simplification, namely, that the $k$th response function can be labelled by the $k$th element of the POVM alone, that is, 
\begin{equation}\label{eq:effectdep}
\xi(k|\lambda,M)=\xi_{E_k}(\lambda).
\end{equation}

The proof is straightforward. Consider a measurement $M'$ that is obtained from $M$ by coarse-graining all outcomes $k\ne k_0$.  Suppose outcome $k_0$ of  $M$ maps to outcome $0$ of $M'$ and any outcome $k\ne k_0$ of $M$ maps to outcome $1$ of $M'$. It is clear that $M'$ is then associated with the two-outcome POVM $\{ E_{k_0}, I-E_{k_0}\}$.  We can therefore label the response functions of $M'$ by this POVM,
\begin{eqnarray}
\xi(j|\lambda,M')=\xi(j|\lambda,\{ E_{k_0}, I-E_{k_0}\}).
\end{eqnarray}
Meanwhile, by the definition of $M'$, 
\begin{eqnarray}
\xi(k_0|\lambda,M)=\xi(0|\lambda,M'),
\end{eqnarray}
and therefore
\begin{eqnarray}
\xi(k_0|\lambda,M)=\xi(0|\lambda,\{ E_{k_0}, I-E_{k_0}\}).
\end{eqnarray}
However, given that the POVM $\{ E_{k_0}, I-E_{k_0}\}$ is completely specified by specifying $E_{k_0}$, it suffices to label the response function by $E_{k_0}$, 
\begin{eqnarray}
\xi(k_0|\lambda,M)=\xi_{E_{k_0}}(\lambda),
\end{eqnarray}
which is what we set out to prove.

It follows from this analysis that in a noncontextual ontological model of operational quantum theory, 
\emph{the response function associated with a given outcome of a quantum measurement depends only on the POVM element associated to that outcome}.

In the rest of this section, we will show how 
our proposal for how to model unsharp measurements in a noncontextual ontological model can be expressed as a set of constraints on probability-assignments to effects, rather than in terms of constraints on response functions.  This manner of expressing the proposal is in some respects easier to grasp and clarifies how it contrasts with alternative proposals.

We begin with the traditional notion of noncontextuality, which, as mentioned in the introduction, applies only to sharp measurements. 
In terms of the notions introduced here, it is the conjunction of the assumption of measurement noncontextuality and the assumption of outcome determinism for all sharp measurements.  Specifically, if $M$ is a sharp measurement with outcomes labelled by $k$ and associated with the projector-valued measure $\{ \Pi_k \}$, then, specializing Eq.~\eqref{eq:effectdep} to projectors, 
measurement noncontextuality implies that
\begin{eqnarray}\label{eq:NCprojector}
\xi(k|\lambda,M)=\xi_{\Pi_k}(\lambda).
\end{eqnarray}
Meanwhile, outcome determinism for sharp measurements implies that
\begin{eqnarray}
\xi_{\Pi_k}(\lambda) \in \{0,1\}.
\end{eqnarray}

The traditional assumption of noncontextuality can also be expressed in terms of constraints on the 0-1 valuation of projectors for a fixed ontic state. 
It is straightforward to verify that this formulation follows from the one in terms of response functions that we have just given.
Let $v(\Pi)$ denote the value assigned to projector $\Pi$ by a fixed ontic state $\lambda$, i.e. $v(\Pi)=\xi_{\Pi}(\lambda)\in \{0,1\}$.  Traditional noncontextuality then 
asserts that for every ontic state, the following conditions hold:
\begin{itemize}
\item[KS1] Each projector $\Pi$ is assigned a value 0 or 1, $v(\Pi)\in \{0,1\}$,
\item[KS2] For each pair of projectors $\Pi_1,\Pi_2$, if $\Pi = \Pi_1 +\Pi_2$ is also a projector, then $v(\Pi)=v(\Pi_1)+v(\Pi_2)$,
\item[KS3] The identity operator is assigned the value 1, $v(I)=1$. 
\end{itemize}
The second item implies that orthogonal projectors cannot both receive the value 1, and the third implies that for any set of projectors that form a resolution of identity, exactly one of them must be assigned the value 1.\footnote{It is tempting to think that constraints KS1-KS3 are the content of the assumption of traditional noncontextuality, but this is inaccurate.  Rather, the assumption of measurement noncontextuality is a prerequisite to KS1-KS3 making any sense. It is this assumption that warrants positing a function $v$ that depends \emph{only} on the projector associated with a measurement outcome.  So once one is discussing the properties of a valuation over projectors, the assumption of noncontextuality has already done its work.  
KS2 then follows from how one must represent coarse-graining in an ontological model (given by Eq.~\eqref{eq:coarsegraining}), KS3 follows from the fact that for every measurement, the sum of probabilities of all the outcomes must be 1, and KS1 encodes the assumption of outcome determinism for sharp measurements.}

We now reconsider the question of how to model unsharp measurements in a noncontextual ontological model from the perspective of valuations over effects. 

The proposal of Refs.~\cite{Busch,Cabello,Aravind, Methot,Renes,Zhang,Mancinska} is that all effects, even the nonprojective ones, should receive 0-1 valuations for a fixed ontic state and that these should satisfy the same constraints as do the 0-1 valuations of projectors:
\begin{itemize}
\item[1] Each effect $E$ is assigned a value 0 or 1, $v(E)\in \{0,1\}$,
\item[2] For each pair of effects $E_1,E_2$, if $E = E_1 +E_2$ is also an effect, then $v(E)=v(E_1)+v(E_2)$,
\item[3] The identity operator is assigned the value 1, $v(I)=1$. 
\end{itemize}
These constraints imply that for every POVM, precisely one effect must receive the value 1 and the others 0.  

In our approach, on the other hand, nonprojective effects are not assigned deterministic values, but only probabilities.  
Nonetheless, one can express our proposal in terms of constraints on the \emph{probability-assignments} to effects.  Recall that an effect $E$ is a  positive operator satisfying $0 \le E \le I$.  Let $w(E)$ denote the probability assigned to effect $E$ by a fixed ontic state.  The constraints on probability assignments to effects are:
\begin{itemize}
\item[NC1] Each effect $E$ is assigned a probability, $w(E)\in [0,1]$,
\item[NC2] For each pair of effects $E_1,E_2$, if $E = E_1 +E_2$ is also an effect, then $w(E)=w(E_1)+w(E_2)$, 
\item[NC3] For each effect $E$ and for any $s$ satisfying $0 \le s \le 1$,  if $sE$ is an effect, then $w(sE)=sw(E)$.
\item[NC4] The identity operator is assigned unit probability, $w(I)=1$. 
\item[NC5] $w(E) \in \{0,1\}$ if and only if the effect $E$ is a projector, i.e. $E^2=E$.
\end{itemize}

We now demonstrate how these constraints can be justified under the assumption of noncontextuality in the sense of definition \ref{def:NC}. 
To translate from `response function' language to `probability assignment' language, we simply note that an effect $E$ describes an operational equivalence class of measurement events and the probability assignment to $E$ for ontic state $\lambda$ is the value of the associated response function at $\lambda$, that is, $w(E) = \xi_E(\lambda)$.  In particular, if outcome $k$ of measurement $M$ is in the equivalence class of measurement events associated with $E$, then $w(E) = \xi(k|M,\lambda)$. 

 NC1 then follows from the fact that $\xi(k|M,\lambda)$ is a probability.  
 
 NC2 is simply a consequence of the representation of coarse-graining in an ontological model.  The equality $E = E_1 +E_2$ implies that $E_1$ and $E_2$ can be associated with two distinct outcomes of a single measurement, and that the coarse-graining of that pair of outcomes is associated with the effect $E$. However, if two outcomes of a measurement are coarse-grained into a single outcome, then the probability for the latter given ontic state $\lambda$ is simply the sum of the probabilities for each of the former given $\lambda$ (see Eq.~\eqref{eq:coarsegraining}).    
 
 NC3 is a consequence of the representation of post-processing in an ontological model.
Suppose the effect $E$ is associated with the measurement event corresponding to implementing the measurement procedure $M$ and obtaining the outcome $k$.
Now define a measurement procedure, $M'$, as a post-processing of $M$ as follows: $M'$ yields outcome $0$ with probability $s$ if $M$ yields outcome $k$. 
Clearly, the effect associated with outcome 0 of measurement $M'$ is then $sE$.  
We have already seen how to represent post-processing in an ontological model (see Eq.~\eqref{eq:postprocessing}).  From this, we infer that $\xi(0|M',\lambda)=s\xi(k|M,\lambda)$, but given that $w(E) = \xi(k|M,\lambda)$ and $w(sE) = \xi(0|M',\lambda)$, we have that $w(sE)=sw(E)$.

NC4 follows from the fact that the probabilities assigned to the outcomes of any measurement must sum to 1.  

Finally, NC5 follows from theorem~\ref{thm:main}.
The only subtlety in making this inference is that theorem~\ref{thm:main}
concerns the representation of the outcomes of sharp measurements and hence the representation of projectors that appear in POVMs \emph{all} of whose elements are projectors
while NC5 refers to a single measurement event associated with a projector, even if that projector appears in a POVM alongside nonprojective effects.  
We bridge the gap by noting that measurement noncontextuality implies that the response function associated with a projector is the same regardless of what measurement that projector is considered a part of (as implied by Eq.~\eqref{eq:NCprojector}).  Because a projector is represented outcome-deterministically when it is part of a sharp measurement (by theorem~\ref{thm:main}),  it must be represented outcome-deterministically even when it is part of an unsharp measurement.  

It is worth noting a few facts about this set of constraints.

First, NC2, NC4 and NC5 together imply KS1, KS2 and KS3, so the usual constraints on how to represent projectors in a noncontextual ontological model are recovered as special cases of our constraints on how to represent arbitrary effects in a noncontextual ontological model. 

It is also useful to note that the constraints NC1-NC5 are equivalent to the following set:
\begin{itemize}
\item[NC1$^\prime$] Each effect $E$ is assigned a probability equal to one of the eigenvalues of $E$, $w(E)\in \textrm{spec}(E)$,
\item[NC2$^\prime$] For each pair of effects $E_1,E_2$ and for each pair of reals $s_1,s_2$ satisfying $0 \le s_1,s_2 \le 1$,  if $s_1 E_1+s_2 E_2$ is an effect, then $w(s_1 E_1+s_2 E_2)=s_1 w(E_1)+s_2 w(E_2)$.
\end{itemize}
Equivalence is easy to prove. It is trivial to see that NC1, NC4 and NC5 all follow from NC1$^\prime$, while NC2 and NC3 follow from NC2$^\prime$.  Conversely, NC2$'$ follows trivially from NC2 and NC3.  Finally, NC1$'$ is derived from NC1-NC5 using the same logic that is used in the proof of theorem~\ref{thm:main}. Any effect $E$ has a spectral resolution of the form $E=\sum_i s_i \Pi_i$ (where $\sum_i \Pi_i = I$) and therefore, by NC2$'$, $w(E)=\sum_i s_i w(\Pi_i)$.  However, NC2, NC4 and NC5 imply that $w(\Pi_i)=1$ for precisely one value of $i$ and is zero otherwise (perhaps the easiest way to see this is by noting that NC2, NC4 and NC5  imply KS1, KS2 and KS3), and this in turn implies that $w(E)=s_i$ for some value of $i$.  Hence, $w(E)\in \textrm{spec}(E)$.

In Sec.~\ref{sec:discussion}, we will show how this manner of characterizing the consequences of noncontextuality for ontological models of quantum theory allows one to simplify existing no-go theorems.  Before doing so, however, we make use of this distinction between 0-1 valuations and probability assignments to frame our final criticisms of ODUM.

\section{A dialogue concerning the status of ODUM in a noncontextual model}\label{sec:coinflipPOVM}

The compelling but incorrect argument for ODUM that we considered in Sec.~\ref{sec:compellingargument} is, in our estimation, the most likely reason that many researchers have assumed ODUM without question.  We hope, therefore, that our critique of this argument is sufficient to convince such researchers to abandon it.  Nonetheless, we have also shown in Sec.~\ref{sec:probassignments} that the assumption of noncontextuality in definition~\ref{def:NC} explicitly implies the failure of ODUM.  In this section, we provide a few more arguments against ODUM.  We present these as a dialogue between the author and an imaginary proponent of ODUM.\footnote{We do not claim that any \emph{actual} proponent of ODUM would make the arguments that are made by our imaginary proponent.  
Nonetheless, certain parts of our dialogue are inspired by various proposals for how to define a notion of noncontextuality for unsharp measurements, as we note explicitly throughout.}
This provides a more direct confrontation between our approach and the proposal 
of Refs.~\cite{Busch,Cabello,Aravind, Methot,Renes,Zhang,Mancinska}.   
Such arguments may seem redundant at this point, but given that the ODUM proposal has been revived several times by various authors, we feel that it may be prudent to drive a few more nails into its coffin.  

We begin by repeating the argument against ODUM that was made in Ref.~\cite{Spe05}. 

\begin{dialogue}

\speak{Author}
Consider a measurement procedure \textrm{M }associated with the
POVM $\{\tfrac{1}{2}I,\tfrac{1}{2}I\}$.  For any quantum state $\rho$, we have $\textrm{tr}(\rho \tfrac{1}{2}I)=\tfrac{1}{2}$, therefore this measurement always has equal probability of producing either of its two outcomes, regardless of the preparation procedure.  Clearly then, one way of implementing this measurement is as follows:
\emph{completely ignore the system} and just flip a fair coin to determine the
outcome. Call this measurement procedure $\mathrm{M}$, and consider how it must be represented in
an ontological model. Because the outcome doesn't depend on the system at all,
it follows that regardless of the system's ontic state $\lambda$, there is a
probability of $1/2$ for each outcome, so it is represented by the set of
response functions $\{\tfrac{1}{2},\tfrac{1}{2}\}$
where each element should be thought of as a uniform function over $\lambda$
of height ${\textstyle{\frac{1}{2}}}$.
The outcomes are clearly not deterministic given $\lambda$, so outcome determinism fails to hold.

\speak{Proponent of ODUM}
Yes, but there are \emph{other} measurement procedures associated
with the POVM $\left\{  \tfrac{1}{2}I,\tfrac{1}{2}I\right\}  $ and one of \emph{these} may be represented by an outcome-deterministic response function.

\speak{Author}
The problem with this response is that measurement noncontextuality (in the sense of definition~\ref{def:NC}) requires that a POVM be represented
by the same set of response functions \emph{regardless} of which particular
measurement procedure is used to implement it.   It therefore suffices
to find just one measurement procedure for the POVM that must be represented outcome-indeterministically to infer that in a noncontextual ontological model, \emph{all}
measurement procedures for this POVM must be represented outcome-indeterministically.

\speak{Proponent of ODUM}
That may be, but it seems to me significant that one can find sets of unsharp measurements, which, unlike the fair coin flip POVM have a nontrivial dependence on the system, and for which 
it is impossible to find a noncontextual and outcome-deterministic representation.
Here's an example of three such POVMs (due to M.\ Nakamura, who was inspired by a similar example due to Cabello, both of which are reported in Ref.~\cite{Cabello}):
\begin{align}
\{ \tfrac{1}{2}E, \tfrac{1}{2}(I-E), \tfrac{1}{2}F, \tfrac{1}{2}(I-F) \}, \nonumber \\ 
\{ \tfrac{1}{2}E, \tfrac{1}{2}(I-E), \tfrac{1}{2}G, \tfrac{1}{2}(I-G) \}, \nonumber \\ 
\{ \tfrac{1}{2}F, \tfrac{1}{2}(I-F), \tfrac{1}{2}G, \tfrac{1}{2}(I-G) \}.
\end{align}
ODUM implies that for a given $\lambda$ every effect
must be assigned a value of 0 or 1, while measurement noncontextuality implies that an effect receives
the same value regardless of where it occurs. 
The contradiction is obtained by noting that only one effect in each POVM can be assigned the value 1, implying an odd number of 1s, but every effect appears in two POVMs so that a noncontextual assignment must assign an even number of 1s.  

Such proofs make use of details of the structure of quantum measurements.  Doesn't their existence show that ODUM is an interesting assumption? 

\speak{Author}
No, they don't.  If we assume ODUM, then the fair coin flip POVM is already sufficient for deriving the contradiction.
Given that the two effects in the fair coin flip POVM are the same, measurement noncontextuality requires that we assign them the same value.  But we can't assign them both the value 0 because this would say that neither outcome occurs, and we can't assign them both the value 1 because this would say that both outcomes occur.  We have our contradiction.

So we see that proofs of the type described above 
are unnecessarily complicated: a consideration of the fair coin flip POVM $\{\tfrac{1}{2}I,\tfrac{1}{2}I\}$ yields the
result immediately.  The fact that the contradiction can be obtained by a completely trivial argument speaks against ODUM.\footnote{Grudka and Kurzynski~\cite{Grudka} have also criticized the notion of noncontextuality used in the Cabello-Nakamura proofs. They argue that in a noncontextual model, one should only assign deterministic values to the projectors that appear in a \emph{Naimark extension} of the POVM, rather than the POVM elements themselves.  It then suffices to note that the projector that extends a given effect varies with the POVM in which that effect appears, and therefore that a noncontextual model does \emph{not} assign a unique deterministic value to a given effect.  In the language of the present article, they argue that a noncontextual and outcome-deterministic value-assignment to projectors on system+ancilla does not imply a noncontextual and outcome-deterministic value-assignment to effects on the system.  This attitude is entirely consistent with the view espoused here.}

\speak{Proponent of ODUM}
Thinking it over, I've refined my view on the matter. \ The problem
isn't with ODUM, the problem is with your definition of measurement noncontextuality. \ It's too
strong. \ The proper notion of measurement noncontextuality for unsharp measurements should demand that equivalent effects
are represented by equivalent response
functions \emph{only when these effects appear in distinct POVMs.}  Measurement noncontextuality should not require equivalent representations for equivalent effects if these appear in the same POVM. In other words, we need only
eliminate context-dependence \emph{between} but not\emph{ within} measurements.

\speak{Author}
In my view, the motivation behind the assumption of measurement noncontextuality is that statistical indistinguishability relative to all preparations should imply statistical indistinguishability relative to all ontic states, therefore events that have the same statistics for all preparations should be represented equivalently in the ontological model, even if they correspond to distinct outcomes of a single measurement.

But in any case, \emph{even if} we consider your suggested modification of the notion of measurement noncontextuality, it is still trivial to obtain
a contradiction. \ Consider the fair coin flip POVM $\{E_{1},E_{2}\}$ where $E_{1}=E_{2}=\tfrac{1}{2}I$
together with another POVM containing $\tfrac{1}{2}I$, say 
$\left\{  F_{1},F_{2},F_{3}\right\}  $ where $F_{1}=\tfrac{1}{2}I$ and $F_{2}=\tfrac{p}{2}I,\;F_{3}=\tfrac{1-p}{2}I$.
According to the notion of measurement noncontextuality that you propose, we must require $F_{1}$ to take the same
value as $E_{1},$ but by the same token we must also require $F_{1}$ to take the same value as
$E_{2}.$  This implies that $E_{1}$ and $E_{2}$ must take the same value,
and so we are back to applying ODUM directly to the fair coin flip POVM and obtaining a contradiction trivially.

\speak{Proponent of ODUM}
Fine, given this example, I propose that the assumption of measurement noncontextuality together with ODUM simply
cannot be applied to any POVM wherein the same effect appears twice.\footnote{This proposal was considered by Methot~\cite{Methot}.}

\speak{Author}
Ah, but this move won't help either; under this new notion one can \emph{still }construct a contradiction essentially trivially. \ Consider the POVM 
$\{\tfrac{p}{2}I,\tfrac{1-p}{2}I,\tfrac{q}{2}I,\tfrac{1-q}{2}I\}$, where $0<p,q<1$, $p,q\ne \tfrac{1}{2}$ and $p \ne q$.
\ Suppose that it is coarse-grained in one of two ways: either one
coarse-grains the first pair of outcomes, in which case the resulting POVM is
$\{\tfrac{1}{2}I,\tfrac{q}{2}I,\tfrac{1-q}{2}I\},$ or one coarse-grains the last pair of outcomes,
in which case the resulting POVM is $\{\tfrac{p}{2}I,\tfrac{1-p}{2}I,\tfrac{1}{2}I\}.$ \ Now recall
that coarse-graining at the operational level is represented by
coarse-graining at the ontological level, that is, if $v(E)$ is the value assigned to effect $E$ by an ontic state $\lambda$, then $v(E_{1}+E_{2})=v(E
_{1})+v(E_{2}).$ Assuming ODUM, precisely one of the four effects in $\{\tfrac{p}{2}I,\tfrac{1-p}{2}I,\tfrac{q}{2}I,\tfrac{1-q}{2}I\}$ must receive the value $1,$ but
then upon coarse-graining, the effect $\tfrac{1}{2}I$ will receive different values
depending on whether it is considered in the context of $\{\tfrac{1}{2}I,\tfrac{q}{2}I,\tfrac{1-q}{2}I\}$ or of
$\{\tfrac{p}{2}I,\tfrac{1-p}{2}I,\tfrac{1}{2}I\}.$ \ For instance, if $v\left( \tfrac{p}{2}I \right)  =1$
while $v(\tfrac{1-p}{2}I)=v(\tfrac{q}{2}I)=v(\tfrac{1-q}{2}I)=0,$ then in the context of the first
coarse-graining, $v(\tfrac{1}{2}I)=v\left(
 \tfrac{p}{2}I \right)  +v\left( \tfrac{1-p}{2}I\right)=1 $,
 while  in the context of the second coarse-graining, $v(\tfrac{1}{2}I)=v\left( \tfrac{q}{2}I\right)  +v\left(
\tfrac{1-q}{2}I\right)=0$.  The same example would of course \emph{not} yield a contradiction if we did not assume ODUM.

\speak{Proponent of ODUM}
You know, I can avoid all of these problems by adopting the following notion of noncontextuality for unsharp measurements: one is only required to assign outcome-deterministic response functions to effects that \emph{cannot} appear more than once in a given POVM, that is, effects $E$ satisfying $E > \tfrac{1}{2}I$.  Given that all of your criticisms of ODUM make use of the effect $\tfrac{1}{2}I$, and this does not fulfill the conditions for applicability of such a notion of noncontextuality, your criticisms would no longer apply.\footnote{This notion was considered by Bacon, Toner and Ben-Or~\cite{TonerBaconBenOr} and reported by Methot~\cite{Methot}.}

\speak{Author}
Isn't this starting to feel like epicycles to you?  In any case, a good operational notion of noncontextuality should apply to \emph{any} measurement.  If some proposed notion of noncontextuality for unsharp measurements necessitates a restriction on the sorts of measurement to which it can be applied, then it hasn't really addressed the problem that needs to be solved.  If one wants to be able to say, of any given experiment, whether it admits of a noncontextual model or not, the definition of noncontextuality must be able to cover all cases.

\speak{Proponent of ODUM}
Well if you insist on a definition that covers all possible measurements, then I'm just going to bite the bullet:
my original idea of assuming ODUM in addition to the general notion of measurement noncontextuality was right all along 
and the impossibility of such an ontological model
is indeed trivial to demonstrate.
If your intuitions suggested that such proofs should be nontrivial, well
then, these examples only demonstrate how wrong those intuitions were.

\speak{Author}
This is a possible position, but it does not have much to recommend it.  Recall that
noncontextuality no-go theorems based on sharp measurements, such as the original Kochen-Specker theorem, make critical use of \emph{structural differences} between the set of quantum measurements and the
set of classical measurements.  All of the trivial no-go theorems I've provided above are based on POVMs wherein every element is proportional to the identity operator.  Every such POVM boils down to sampling from a probability distribution, yielding outcome statistics that are completely independent of the preparation procedure.  But of course a \emph{classical} operational theory also admits of noisy measurements that just correspond to sampling from a probability distribution and yield outcome statistics that are independent of the preparation procedure.  Therefore, these trivial no-go theorems do not rely on
any intrinsically quantum features of the measurements.

\speak{Proponent of ODUM}
Maybe what this shows is that we can obtain a no-go result for
noncontextuality even for \emph{classical} operational theories!

\speak{Author}
I take it to be a point in favour of 
\emph{not} assuming ODUM that ontological models of \emph{classical} operational theories are always found
to be noncontextual in the sense of definition~\ref{def:NC}.  The whole point of these kinds of investigations is to identify the ways in which quantum theories and classical theories differ, so a good notion of noncontextuality is one that can do justice to the difference.

\speak{Proponent of ODUM}
Well, in the end, it's just a question of semantics what one decides to
call \textquotedblleft noncontextual\textquotedblright. \ I want to give the
name to one kind of model, you want to give it to another.

\speak{Author}
I don't agree that the debate is just about semantics. There \emph{are}
criteria by which we can judge different proposed generalizations of the
notion of noncontextuality: coherence and usefulness. 

 This article has sought to demonstrate that the generalized notion of measurement noncontextuality proposed in Ref.~\cite{Spe05}, wherein unsharp measurements are associated with outcome-indeterministic response functions, is coherent.  It reduces to the traditional notion of noncontextuality for sharp measurements and does justice to the manner in which unsharp measurements can be related to sharp measurements, for instance, when one is a post-processing of the other.
 
The proposed notion of noncontextuality also fulfils the criterion of usefulness insofar as
it allows us to clarify the ways in which classical and quantum theories
differ. Indeed, certain information-processing tasks can be implemented better in operational theories that
do not admit of an ontological model that is noncontextual in the sense of definition~\ref{def:NC}, such as the task of parity-oblivious multiplexing described in Ref.~\cite{ParityObliviousMulitplexing}. \ A definition of
noncontextuality that does not cleanly distinguish between classical
operational theories and quantum operational theories is less useful than one
that does.

\direct{End scene.}
\end{dialogue}

\section{Discussion}
\label{sec:discussion}

This article has focussed on the question of how to formulate the notion of noncontextuality for unsharp measurements.  This informs the question of whether one can rule out a noncontextual model of quantum theory using a proof that appeals explicitly to unsharp measurements.  Although there are several results in the literature that claim to have done precisely this~\cite{Busch,Cabello,Renes,Aravind,Methot}, these have all assumed ODUM.  This assumption does not follow from an assumption of noncontextuality, however, and consequently in the face of a contradiction, one can choose to abandon ODUM in order to salvage noncontextuality.  In our view, therefore, these proofs do not deserve to be called no-go theorems for noncontextuality.

Nonetheless, it \emph{is} possible to construct a no-go theorem for noncontextuality for a single qubit without the assumption of ODUM: this was done in Sec. V of Ref.~\cite{Spe05}, where two such proofs are presented, one based on a finite set of measurements on a qubit and another that is based on a version of Gleason's theorem for unsharp measurements that applies to two-dimensional Hilbert spaces, proven by Busch~\cite{Busch} and by Caves {\it et al.}~\cite{Renes}.

We repeat these proofs here, using the characterization of noncontextuality in terms of probability assignments to effects (NC1$'$ and NC2$'$ in Sec.~\ref{sec:probassignments}), which simplifies them substantially.

First, the discrete proof.
Consider the trine POVM $\{\tfrac{2}{3}\Pi_1,\tfrac{2}{3}\Pi_2,\tfrac{2}{3}\Pi_3\}$, where $\Pi_1,\Pi_2$ and $\Pi_3$ are the rank-1 projectors which in the Bloch-sphere representation correspond to three vectors in a plane separated from one another by angles of 120$^{\circ}$, so that
\begin{equation}
\tfrac{2}{3}\Pi_1+\tfrac{2}{3}\Pi_2+\tfrac{2}{3}\Pi_3 =I.
\end{equation}
The assumption of a noncontextual model (via NC2$'$) implies that the probability assignment over effects induced by an ontic state must satisfy
\begin{equation}\label{eq:POVMrelation}
\tfrac{2}{3}w(\Pi_1)+\tfrac{2}{3}w(\Pi_2)+\tfrac{2}{3}w(\Pi_3) =w(I).
\end{equation}
However, the assumption of a noncontextual model (via NC1$'$) also implies that
$w(I)=1$ while each of $w(\Pi_1)$, $w(\Pi_2)$ and $w(\Pi_3)$ must take the value 0 or 1.  However, no such assignment in consistent with Eq.~\eqref{eq:POVMrelation}, so we have derived a contradiction.  Note that this proof makes use of the distinctive features of the space of quantum effects.

The Gleason-like proof is also straightforward. 
NC2$'$ implies that any probability assignment $w$ to the effects is convex-linear in the effects.  By a standard argument~\cite{Busch,Renes}, it can therefore be extended to a linear function over Hermitian operators and hence represented as the inner product with another Hermitian operator.  Denoting this operator by $\rho$, and recalling that the Hilbert-Schmidt inner product between Hermitian operators $A$ and $B$ is $\textrm{tr}(AB)$, we have  $w(E)=\textrm{tr}(\rho E)$.  The fact that $w(E)$ is required to be positive for all $E$ implies that $\rho$ must be a positive operator, and the fact that we require $w(I)=1$ implies that $\textrm{tr}(\rho)=1$, therefore $\rho$ is a density operator.  To get a contradiction, it then suffices to note that there is no density operator that assigns to every effect a value from its spectrum. In particular, there is no quantum state that assigns values 0 or 1 to every projector.

Another single-qubit no-go theorem for noncontextuality has been provided recently by Kunjwal and Ghosh~\cite{KunjwalGhosh} who demonstrate a violation of a noncontextuality inequality derived in Ref.~\cite{LiangSpekkensWiseman} for a triple of nonprojective qubit POVMs that are jointly measurable pairwise but not triplewise. 

One consequence of our analysis is that the restriction of previous no-go theorems for noncontextuality to Hilbert spaces of dimension 3 or greater was an artifact of having a notion of noncontextuality that was limited to sharp measurements.  For a qubit, there is only a single measurement context in which any given rank-1 projector can appear, namely, together with its unique rank-1 orthogonal complement.  Hence, there is no possibility of a nontrivial variation of the context in which a projector appears and hence no possibility of context-dependence either.  When one considers unsharp measurements, on the other hand, there \emph{are} nontrivial contexts: a given nonprojective POVM may be realized as a convex combination of other measurements in multiple ways, as a post-processing of other measurements in multiple ways, and by reduction of another measurement in multiple ways.  However, for unsharp measurements, achieving a noncontextual model is not about assigning \emph{outcomes} in a context-independent fashion, it is about assigning \emph{probabilities} of outcomes in a context-independent fashion.

Finally, a few words are in order regarding the motivation that we provided for our investigation---determining whether experimental statistics can be explained by a noncontextual ontological model, irrespective of the truth of quantum theory.
This article has not addressed this question directly; 
we have only considered the constraints on noncontextual ontological models \emph{of quantum theory}.  The distinction between sharp and unsharp measurements, for instance, is defined in terms of quantum concepts. What our analysis demonstrates, however, is that one should never simply \emph{assume} that the noncontextual representation of some measurement should be outcome-deterministic.  Rather, this needs to be justified.  As we have noted, Ref.~\cite{Spe05} demonstrated that one can justify outcome-deterministic representations of sharp measurements in quantum theory from preparation noncontextuality and from certain facts about the quantum statistics.   In the context of exploring whether given experimental data can be explained by a noncontextual ontological model, any assumptions of outcome-determinism will similarly need to be justified by assumptions of noncontextuality and by appeal to facts about the experimental statistics.  Or, more precisely, the \emph{degree} of indeterminism posited for some noncontextual representation of a measurement needs to be so justified.   An example of a noncontextuality inequality for experimental statistics wherein the degree of indeterminism is justified in this manner will be provided elsewhere~\cite{SpekKunj}.

\begin{acknowledgements}
The author would like to thank John Sipe and Howard Wiseman for their insistence that the topic of outcome-determinism for unsharp measurements deserved a better treatment.  Thanks also to Ernesto Galv\~{a}o, Ben Toner, Howard Wiseman, and Ravi Kunjwal for discussions, and to an anonymous referee for suggesting a simplification of the proof of Thm.~\ref{thm:main}.  Research at Perimeter Institute is supported in part by the Government of Canada through NSERC and by the Province of Ontario through MRI.
\end{acknowledgements}

\appendix

\section{Ontological extension: modelling an unsharp measurement using an
outcome-deterministic response function}\label{app:ontext}

The intuition at play in premiss 2 of the argument from Sec.~\ref{sec:compellingargument} is that one can always imagine any
subjective uncertainty in the outcome of a measurement on some system $s$ as
being due to uncertainty about the ontic state of some other
system that, together with $s,$ determines the outcome, for instance, hidden
variables in the apparatus.   We will see that this is indeed the case.  The
resulting representation of the measurement will be called an \emph{ontological extension}.  This is the analogue, within the ontological model, of 
the Naimark extension of a measurement within operational quantum theory.~\footnote{Just as one can define a quantum Naimark
extension by adjoining an ancilla to one's system or by considering the system
to be a subspace of a larger system, so too can one define an ontological
Naimark extension in either way. We'll use the ancilla construction here.
It is possible that one could dispense with the assumption of separability if one used the subspace
construction, but we do not seek to answer the question here.}

Suppose that for an operational theory $\left(  \mathcal{P},\mathcal{M}%
,p\right)  $ on system $s,$ we have found a (possibly contextual) ontological
model $\left(  \Lambda_{s},\mu,\xi\right)  $ on $s,$ such that the
operational statistics are reproduced as%
\begin{equation}
p(k|M,P)=\sum_{\lambda_{s}\in\Lambda_{s}}\mu(\lambda_{s}|P)\xi(k|\lambda_{s},M).
\end{equation}
We suppose that the ontological model is outcome-indeterministic, so that
$0<\xi(k|\lambda_{s},M)<1$ for some $\lambda_{s}.$ \ 

We can define a new model with outcome-deterministic response functions as
follows. \ Introduce an ancilla system $a$ with ontic state space equal to the unit interval (hence continuous), that is, $\Lambda
_{a}=[0,1]$. The ontic state space of the composite is
then $\Lambda_{sa}\equiv\Lambda_{s}\times\Lambda_{a}.$ \ Now define%

\begin{equation}
\mu(\lambda_{s},\lambda_{a})\equiv\mu(\lambda_{s}|P)\mu_{0}(\lambda
_{a}),
\end{equation}
where $\mu_{0}(\lambda_{a})$ is the uniform distribution over $\left[
0,1\right]  $. \ Also, define%
\begin{align}
\omega_{0}(\lambda_{s})&=0, \\
\omega_{k}(\lambda_{s})&=\sum_{j=1}^{k}%
\xi(j|\lambda_{s},M),
\end{align}
and define outcome-deterministic response functions on $\Lambda_{s}%
\times\Lambda_{a}$ by%
\begin{equation}
\xi\left( k| \lambda_{s},\lambda_{a}\right)  \equiv \left\{
\begin{tabular}
[c]{l}
$1\text{ if }\omega_{k-1}(\lambda_{s})\leq\lambda_{a}\leq\omega_{k}%
(\lambda_{s})$\\
$0$ otherwise
\end{tabular}
\; \; .\right.
\end{equation}
The new model is empirically adequate because
\begin{align}
&  \sum_{\lambda_{s}\in\Lambda_{s}}\int_{\Lambda_{a}}\textrm{d}\lambda_{a}\; \mu(\lambda_{s},\lambda_{a})\xi(k|\lambda_{s},\lambda_{a})\\
&  =\sum_{\lambda_{s}\in\Lambda_{s}}\mu(\lambda_{s}|P) 
\int_{\omega_{k-1}(\lambda_{s})}^{\omega_{k}(\lambda_{s})}\textrm{d}\lambda_{a} \; \mu_{0}\left( \lambda_{a}\right) \\
&  =\sum_{\lambda_{s}\in\Lambda_{s}}\mu(\lambda_{s}|P) \left[  \omega
_{k}(\lambda_{s})-\omega_{k-1}(\lambda_{s})\right] \\
&  =\sum_{\lambda_{s}\in\Lambda_{s}}\mu(\lambda_{s}|P)\xi(k|\lambda_{s},M)\\
&  =p\left(  k|P,M\right)  .
\end{align}

A similar trick for devising an ontological model that is
outcome-deterministic was described by Bell~\cite{Bell} and its relevance for
eliminating determinism as an assumption in the proof of Bell's theorem was
emphasized by Fine~\cite{Fine}.
It follows that one can always
extend an ontological model such that it represents measurements outcome-deterministically if one so wishes.  This justifies premiss $2'$ of the second argument in Sec.~\ref{sec:compellingargument}.

\section{The multiplicity of Naimark extensions of a POVM and its significance for ontological models}\label{app:Naimark}

We here show explicitly that a given POVM can have two distinct Naimark extensions, that is, that there are multiple
distinct choices of projective measurement on $sa$ that yield the same POVM on $s.$
Specifically, we construct two Naimark extensions of the fair coin flip POVM $\{ \tfrac{1}{2} I, \tfrac{1}{2} I\}$ (discussed in Sec.~\ref{sec:coinflipPOVM}).

The first Naimark extension is as follows. We implement a preparation of the
ancilla corresponding to the mixed state%
\[
\rho_a=\tfrac{1}{3}\left\vert 1\right\rangle \left\langle
1\right\vert +\tfrac{1}{3}\left\vert 2\right\rangle \left\langle 2\right\vert
+\tfrac{1}{3}\left\vert 3\right\rangle \left\langle 3\right\vert .
\]
The measurement on $sa$ is the binary-outcome projector-valued measure $\{ \Pi_{sa}, I-\Pi_{sa} \}$, where
\begin{align}
\Pi_{sa} &\equiv  \Pi_{s,1}\otimes\left\vert1\right\rangle \left\langle 1\right\vert
+\Pi_{s,2}\otimes\left\vert 2\right\rangle \left\langle 2\right\vert +\Pi
_{s,3}\otimes\left\vert 3\right\rangle \left\langle 3\right\vert .
\end{align}
where $\Pi_{s,i}\equiv \left\vert\psi_i \right\rangle \left\langle \psi_i\right\vert$ and $\{ \left\vert\psi_i \right\rangle : i \in \{1,2,3\}\}$ are a triple pure states on the system $s$ with pairwise overlaps $|\langle \psi_i | \psi_j \rangle|^2 = \tfrac{1}{4}$ for $i\ne j$, that is, in the Bloch representation, they are separated by $120^{\circ}$ on an equatorial plane of the Bloch sphere. To see that the effective measurement on the system is the POVM
$\{\frac{1}{2} I,\frac{1}{2} I\}$, it suffices to note that
\begin{align*}
&  \mathrm{Tr}_a\left(  \Pi_{sa} \rho_a \right)  =\tfrac{1}{3}\Pi_{s,1}+\tfrac{1}{3}\Pi_{s,2}+\tfrac{1}{3}\Pi_{s,3}  =\tfrac{1}{2}I.
\end{align*}

The second Naimark extension is as follows. We implement a preparation of the
ancilla corresponding to the mixed state
\begin{equation}
\sigma_a=\tfrac{1}{2}\left\vert 1\right\rangle \left\langle 1\right\vert +\tfrac{1}{2}\left\vert 2\right\rangle \left\langle 2\right\vert,
\end{equation}
and implement a measurement on the composite of system+ancilla corresponding
to the three-outcome projector-valued measure
\begin{align}
&  \{I\otimes\left\vert 1\right\rangle \left\langle 1\right\vert
,I\otimes\left\vert 2\right\rangle \left\langle 2\right\vert ,I\otimes
\left\vert 3\right\rangle \left\langle 3\right\vert \}\\
&  \equiv\{\Pi_{sa,1},\Pi_{sa,2},\Pi_{sa,3}\}
\end{align}
(although technically this is a measurement on the composite, it is obviously
only nontrivial on the ancilla). \ Again, it is straightforward to verify that
the effective measurement on the system is the POVM $\{\tfrac{1}{2}I,\tfrac{1}{2}I\}$.

The pair of projector-valued measures appearing in the two Naimark extensions, $\left\{  \Pi_{sa},I-\Pi_{sa}\right\}  $ and $\{\Pi_{sa,1},\Pi_{sa,2},\Pi_{sa,3}\}$, are clearly distinct.  This implies that there are quantum states on $sa$ that yield different statistics for the two.
For instance, the state $\tfrac{1}{2}I\otimes\left\vert 3\right\rangle \left\langle
3\right\vert $ yields 50/50 statistics for $\left\{  \Pi_{sa},I-\Pi_{sa}\right\}  $ but
always yields the third outcome for  $\{\Pi_{sa,1},\Pi_{sa,2},\Pi_{sa,3}\}$.  This implies that although these projector-valued measures are each represented by a set of \emph{outcome-deterministic} response functions in an ontological model that is noncontextual in the sense of definition~\ref{def:NC} (by virtue of
theorem~\ref{thm:main}), nonetheless the two sets of response functions \emph{must be different} to account for the
differing statistics.

\section{An explicit example of ontological extension}\label{app:exampleontext}

We have seen in appendix~\ref{app:Naimark} that because a quantum measurement can be Naimark-extended
in many ways, and because those extensions may be statistically distinguishable, it
follows that the sets of response functions that represent these Naimark extensions must also be inequivalent. 
This section provides a second way of understanding the fact that the cost of modelling statistically-indistinguishable measurements outcome-deterministically is that on the extended system, their representations are no longer equivalent. 
We consider the trick described in appendix~\ref{app:ontext} for replacing a response function on the system with an outcome-deterministic response function on an extension of the system.  Specifically, we show by example that a pair of measurements that are represented by the \emph{same} set of response functions on the system may need to be represented by \emph{different} sets of response functions on its extension.

Let $M$ be a measurement associated with the fair coin flip POVM $\{ \tfrac{1}{2} I, \tfrac{1}{2} I\}$ that was discussed in Sec.~\ref{sec:coinflipPOVM}.
Let the second measurement $M^{\prime}$ be defined as follows.
\begin{quote}
The procedure $M'$: implement $M$, and upon
obtaining outcome $b\in\left\{  0,1\right\}  ,$ output $b\oplus1$, that is, the bit-flip of $b$.
\end{quote}
Equivalently, we can say simply that $M'$ is a measurement of $M$ with the outcomes permuted.  Clearly
$M^{\prime}$ is also represented by the fair coin flip POVM.
Now consider how to represent $M$ and $M^{\prime}$ within a noncontextual
ontological model. As argued in Sec.~\ref{sec:coinflipPOVM}, the set of response functions must be simply $\{ \tfrac{1}{2},\tfrac{1}{2}\}$.\footnote{As an aside, note that the pair $M$ and $M'$ provide \emph{yet another way} of understanding why the POVM $\{ \tfrac{1}{2} I, \tfrac{1}{2} I\}$ must be represented by the set of response functions  $\{ \tfrac{1}{2},\tfrac{1}{2}\}$.  If the set of response functions representing $M$ is denoted $\left\{  \xi\left( 0| \lambda, M\right)  ,\xi\left(1|  \lambda,M \right)  \right\}  $, and similarly for $M'$, then
from the definition of $M^{\prime}$, we must have
\begin{align}
\xi \left( 0|  \lambda,M'\right)   &  =\xi\left(  1| \lambda,M \right)  ,\\
\xi \left( 1|  \lambda,M'\right)   &  =\xi\left(  0| \lambda,M \right)  .
\end{align}
Because $M$ and $M'$ are statistically indistinguishable for all preparations, by measurement noncontextuality they must be
represented by the same set of response functions.  Hence
\begin{align}
\xi \left( 0|  \lambda,M'\right)   &  =\xi\left(  0| \lambda,M \right)  ,\\
\xi \left( 1|  \lambda,M'\right)   &  =\xi\left(  1| \lambda,M \right)  .
\end{align}
But this implies that $\xi\left(  0| \lambda,M \right) =\xi\left(  1| \lambda,M \right)$, and given that $\xi\left(  0| \lambda,M \right) +\xi\left(  1| \lambda,M \right) =1$ for all $\lambda\in\Lambda,$ it follows that
\begin{equation}
\xi\left(  0| \lambda,M \right) =\xi\left(  1| \lambda,M \right) =\tfrac{1}{2}\;\forall\lambda\in\Lambda.
\end{equation}
}

By the construction in appendix~\ref{app:ontext}, we can model both $M$ and $M'$ on a larger system, a composite of system $s$ and an ancilla $a$ with ontic state space $\Lambda_s \times \Lambda_a$, in such a way that the response functions are outcome-deterministic.  Specifically, if we apply this construction to $M$, we obtain
\begin{align}
\xi\left( 0|  \lambda_{s},\lambda_{a},M \right)   &  \equiv \left\{
\begin{tabular}
[c]{l}%
$1\text{ if }0\leq\lambda_{a}\leq1/2$\\
$0$ otherwise
\end{tabular}
\ ,\right. \\
\xi\left(1|  \lambda_{s},\lambda_{a},M\right)   &  \equiv \left\{
\begin{tabular}
[c]{l}%
$1\text{ if }1/2\leq\lambda_{a}\leq1$\\
$0$ otherwise
\end{tabular}
\ .\right.
\end{align}
Meanwhile, if we remember the definition of $M^{\prime},$ it is clear that it
must be represented by a set of response functions that is simply the
permutation of those for $M$,
\begin{align}
\xi\left( 0| \lambda_{s},\lambda_{a},M'\right)   &  =\xi\left(  1|\lambda_{s},\lambda_{a},M\right),\\
\xi\left(1|  \lambda_{s},\lambda_{a},M'\right)   &  =\xi\left( 0| \lambda_{s},\lambda_{a},M\right).
\end{align}
Now we see that although we've managed to have both $M$ and $M^{\prime}$
represented by sets of response functions on $\Lambda_s \times \Lambda_a$ that are outcome-deterministic, the response functions are not equivalent,
\begin{align}
\xi\left( 0| \lambda_{s},\lambda_{a},M'\right)   &  \ne \xi\left(  0|\lambda_{s},\lambda_{a},M\right),\\
\xi\left(1|  \lambda_{s},\lambda_{a},M'\right)   &  \ne \xi\left( 1| \lambda_{s},\lambda_{a},M\right) .
\end{align}

\end{document}